\documentclass[twocolumn,aps,prl,preprintnumbers]{revtex4-2}
\usepackage{bbm}
\usepackage[latin9]{inputenc}
\usepackage{float}
\usepackage{amsmath}
\usepackage{amssymb}
\usepackage{graphicx}
\usepackage{mathrsfs}
\usepackage{amsfonts}
\usepackage{amsthm}
\usepackage{color}
\usepackage{txfonts}
\usepackage[colorlinks=true,citecolor=blue,linkcolor=blue,urlcolor=blue,anchorcolor=blue]{hyperref}%
\usepackage{pifont}
\hypersetup{colorlinks=true,citecolor=blue,linkcolor=blue,urlcolor=blue}
\providecommand{\U}[1]{\protect\rule{.1in}{.1in}}
\makeatletter
\@ifundefined{textcolor}{}
{
\definecolor{BLACK}{gray}{0}
\definecolor{WHITE}{gray}{1}
\definecolor{RED}{rgb}{1,0,0}
\definecolor{GREEN}{rgb}{0,1,0}
\definecolor{BLUE}{rgb}{0,0,1}
\definecolor{CYAN}{cmyk}{1,0,0,0}
\definecolor{MAGENTA}{cmyk}{0,1,0,0}
\definecolor{YELLOW}{cmyk}{0,0,1,0}
}

\begin{document}
\title{Magnon-Driven Phononic Frequency Comb in Linear Elastic Media}
\author{Ziyang Yu\textsuperscript{1,2}}
\author{Zhejunyu Jin\textsuperscript{1}}
\author{Qianjun Zheng\textsuperscript{1}}
\author{Peng Yan\textsuperscript{1,3}}
\email{Contact author: yan@uestc.edu.cn}

\affiliation{\textsuperscript{1}School of Physics and State Key Laboratory of Electronic Thin Films and Integrated Devices, University of Electronic Science and Technology of China, Chengdu 610054, China\\ \textsuperscript{2}Hubei Key Laboratory of Optical Information and Pattern Recognition, 
School of Optical Information and Energy Engineering, Wuhan Institute of Technology, Wuhan 430205, China\\
\textsuperscript{3}Institute of Fundamental and Frontier Sciences, Key Laboratory of Quantum Physics and Photonic Quantum Information of the Ministry of Education, University of Electronic Science and Technology of China, Chengdu 611731, China}

\begin{abstract}
Phononic frequency combs (PFCs) typically require nonlinear elastic media, limiting their frequency range and stability. Here, we propose a transformative approach to generate PFCs in purely linear elastic media by harnessing the magnon nonlinearities, offering a new paradigm for frequency comb physics. By tuning the magnon-phonon coupling confined in a magnetic disk of a vortex state into the strong coupling regime, we demonstrate an efficient nonlinearity transfer from magnons to phonons. This mechanism is able to produce GHz-range PFCs with comb spacing set by the vortex core's gyration frequency. Full micromagnetic simulations verify our theoretical predictions, confirming robust comb formation at 3.5 GHz with 0.4 GHz spacing. This approach overcomes the sub-MHz constraints of conventional PFCs, enabling applications in high-precision metrology, nanoscale sensing, and quantum technologies. Our findings also deepen the understanding of the nonlinear dynamics in hybrid magnon-phonon systems and provide a versatile platform for exploring frequency combs in diverse physical systems.

\end{abstract}

\maketitle
Frequency combs, characterized by equally spaced spectral lines, have transformed precision metrology since their development in optics \cite{newman2019architecture, suh2018soliton, trocha2018ultrafast, wang2020long, liang2015high, lucas2020ultralow, liu2020photonic, marin2017microresonator, corcoran2020ultra, wang2020quantum}. Inspired by the success of optical frequency combs \cite{jones2000carrier, holzwarth2000optical, udem2002optical, herr2016dissipative, zhang2019broadband}, analogous phenomena have been explored in acoustic and magnetic systems, enabling advanced sensing and information processing \cite{cao2014phononic, wang2021magnonic}. Phononic frequency combs (PFCs), in particular, are an exciting frontier in micromechanics and optomechanics \cite{jong2023mechanical, ganesan2017phononic, miri2018optomechanical, hu2021optomechanical}, offering precise frequency references for mechanical systems. However, the PFC generation typically relies on nonlinear media, where challenges like phase mismatches and high amplitude thresholds persist. In addition, dual modes must often be driven at frequencies beyond dispersion bands, with amplitudes exceeding thresholds to enable mode coupling and counter damping \cite{miri2018optomechanical, PhysRevApplied.19.014031, 10.1063/5.0025314}. These intense drives may destabilize systems into chaotic regimes, and the weak nonlinearity and high dissipation of phonons usually restrict PFCs to the sub-MHz range, limiting their high-frequency applications \cite{PhysRevApplied.5.034002, D0EE03014G}.

Magnons, the quanta of spin waves, excel in coherent information processing due to their long propagation lengths and lack of Joule heating \cite{chai2022singlesideband, shen2022coherent, wang2018bistability, korber2023modification, wang2022twisted, yuan2022}. When coupled with phonons, they form magnon polarons---hybrid quasiparticles blending magnon and phonon traits \cite{PhysRevB.91.104409}---extensively studied experimentally \cite{zhang2016cavity, shen2022mechanical, potts2021dynamical, xu2023magnonic, PhysRevB.102.144438, PhysRevLett.117.207203,Shi2021,Hayashi2018,Junxue2020} and theoretically \cite{li2018entanglement, xiong2023frequency}.  It has been shown that the magnon-phonon coupling can drive ferromagnetic resonance in the radio-frequency regime \cite{weiler2012spin} and enables effects like parametric amplification, bistability, and dynamical backaction in materials such as yttrium iron garnet (YIG) \cite{zhang2016cavity, shen2022mechanical, potts2021dynamical}. Recently, acoustic-wave-driven magnonic frequency combs (MFCs) have been realized \cite{xiong2023frequency, xu2023magnonic}, mirroring progress of optomechanical combs  \cite{wang2024optomechanical} in nonlinear elastic media. Yet, it remains uncharted whether magnons can serve as nonlinear actuators to generate PFCs in purely linear elastic media, potentially sidestepping the stringent thresholds of traditional methods.

\begin{figure}
    \centering
    \includegraphics[width=0.48\textwidth]{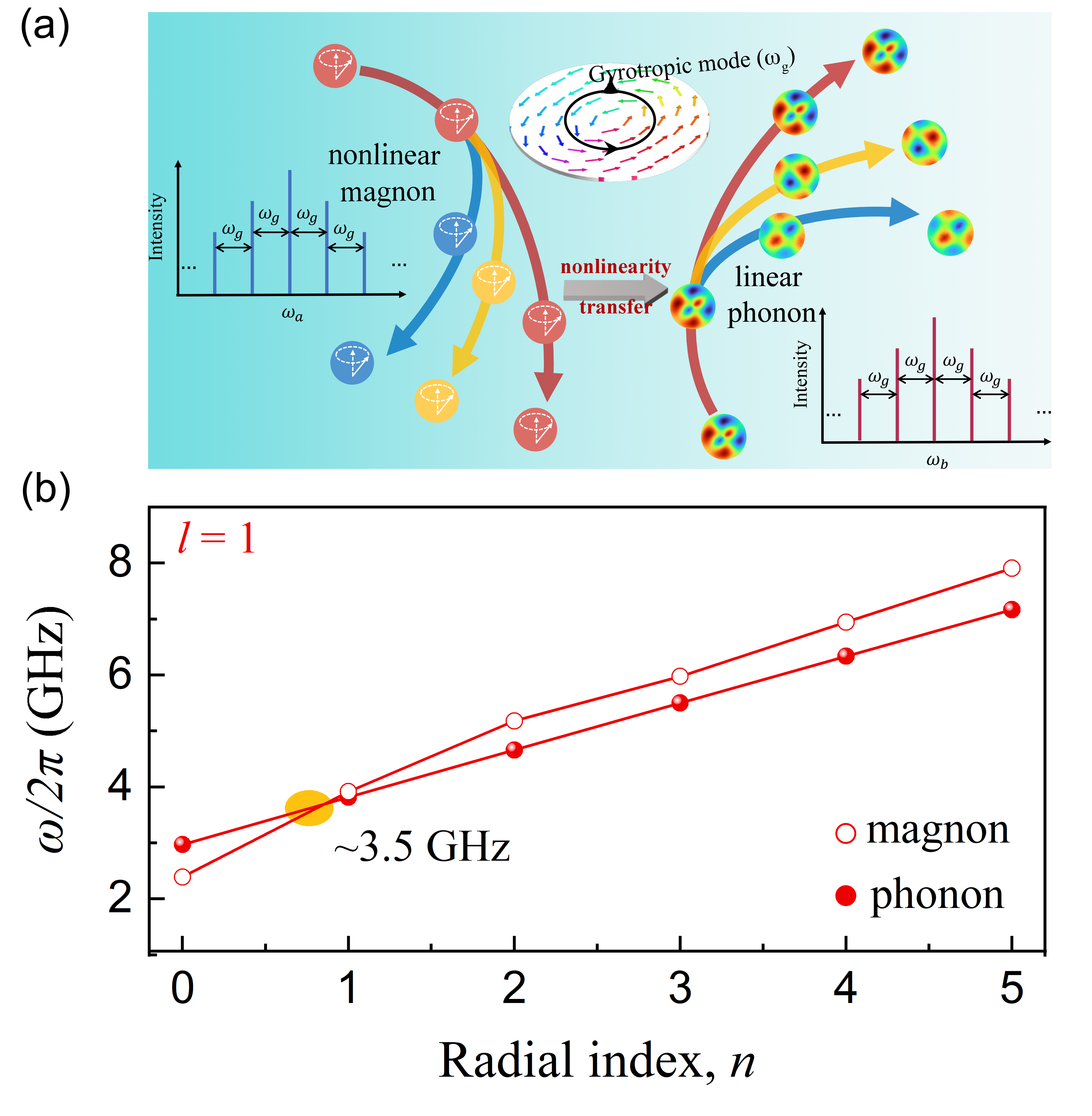}
    \caption{\label{fig1} 
    (a)  Schematic of the nonlinearity transfer from magnons to phonons in the strong-coupling regime, with the comb spacing set by the vortex core gyration frequency (\(\omega_g\)).
    (b) Dispersion relations for magnon and phonon modes with orbital angular momentum quantum number \(l=1\), intersecting near 3.5 GHz, where a strong magnon-phonon coupling may occur (highlighted in yellow).}
\end{figure}

In this Letter, we introduce a novel mechanism for generating PFCs in purely linear elastic media by exploiting magnon-phonon interactions within the strong-coupling regime. Our approach harnesses the intrinsic nonlinearity of magnons to drive the formation of phonon combs, eliminating the need for nonlinear elastic media. This shift not only circumvents the stringent requirements of conventional methods but also offers a fresh and innovative perspective on PFC generation, leveraging the interplay between magnetic and elastic degrees of freedom. To demonstrate this mechanism, we employ a magnetic disk in a vortex state as our model system \cite{yu2021magnetic}. This choice is informed by recent advances in the field, including theoretical studies \cite{10.1063/5.0168968} and experimental work \cite{Vadym2024,Seeger2025} that have investigated the excitation of vortex gyrotropic modes using surface acoustic waves, as well as the realization of MFC \cite{heins2025controlmagnonfrequencycombs, korber2020nonlocal, korber2023modification}. We reveal that when the dispersion of magnons and phonons intersect, the strong coupling enables the transfer of the inherent magnon nonlinearity to phonons, which effectively induces a nonlinear response in the linear phonon system. As a result, a distinct comb-like structure emerges in the spectra of both magnons and phonons, with characteristic frequencies reaching the GHz range---a significant leap beyond the sub-MHz constraints of traditional PFCs. A defining feature of our approach is the precise control over the comb's spectral spacing, which is set by the gyration frequency of the vortex core, as illustrated in Fig.~\ref{fig1}(a). This direct relationship between the magnetic vortex dynamics and the resulting phonon comb properties underscores the elegance of using magnon nonlinearities within a linear elastic framework. This work fundamentally redefines the paradigm of PFC generation, showcasing how magnon can serve as a nonlinear driver in otherwise linear elastic media to produce robust and high-frequency combs.

We begin by constructing the Hamiltonian that governs the system's dynamics
\begin{equation}
H =H_{\text{m}}+H_{\text{p}}+H_{\text{I}},
\end{equation}with \( H_{\text{m,p,I}} \) representing the Hamiltonian of the magnon (\( m \)), phonon (\( p \)) subsystems and their interaction (\( I \)). For nonlinear magnon--vortex interactions in ferromagnetic nanodisks, the minimal model of the magnonic Hamiltonian can be expressed as \cite{wang2022twisted}
\begin{align}
H_{\text{m}} &= \omega_l a_l^{\dagger} a_l + \omega_g a_g^{\dagger} a_g + \omega_p a_p^{\dagger} a_p + \omega_q a_q^{\dagger} a_q \notag \\
  &\quad + \underbrace{g_p \left( a_l a_g a_p^{\dagger} + \text{H.c.} \right) + g_q \left( a_l a_g^{\dagger} a_q^{\dagger} + \text{H.c.} \right)}_{H_{\mathrm{NL}}} \notag \\
  &\quad + \gamma h \left( a_l e^{i \omega_0 t} + \text{H.c.} \right),\label{totalh}
\end{align}
where \( \omega_l \) is the incident magnon frequency, \( \omega_g=5\gamma\mu_{0}M_{s}d/(9\pi R) \) is the vortex core gyration frequency \cite{guslienko2006} with $\gamma$ the gyromagnetic ratio, $\mu_0$ the vacuum permeability, and $d$ and $R$ being the thickness and radius of the nanodisk, respectively. \( \omega_{p,q}=\omega_l \pm\omega_g \) are the sum- and difference-frequencies. The term \(H_{\mathrm{NL}} \) describes nonlinear three-magnon processes, with \( g_p \) and \( g_q \) as the confluence and splitting interaction strength, respectively. The external microwave field has amplitude \( \ h \) and frequency \( \omega_0 \approx\omega_l\). \( a_{l,g,p,q} \)\ (\( a_{l,g,p,q}^{\dagger} \)) are annihilation (creation) operators for the corresponding modes.

The phonon Hamiltonian is
\begin{equation}
H_{\text{p}} = \omega_{\mathit{nl}} b^\dagger b,
\end{equation}
where \(\omega_{\mathit{nl}}= {\gamma _{\mathit{nl}} c_T}/{R}\) is the phonon frequency \cite{landau1986theory, achenbach2012wave} and \(b\ (b^\dagger)\) is the phonon annihilation (creation) operator. Here \( \mathit{n} \) and \( \mathit{l} \) are two integers denoting the radial and orbital angular momentum (OAM) quantum numbers, respectively, \( \gamma_{\mathit{nl}} \) is the \( n \)-th zero of the \( l \)-th order Bessel function of the first kind \( J_l(\gamma_{\mathit{nl}}) = 0 \), \(c_T = \sqrt{\mu/\rho}\) is the transverse phonon velocity, \(\mu\) is the shear modulus, and \(\rho\) is the mass density.

The resonant magnon-phonon interaction is (see Supplemental Material \cite{sup})
\begin{equation}
H_\text{I} =  g_{mp} \left( a_l b^\dagger + a_l^\dagger b \right),
\end{equation}
with the coupling strength \( g_{mp}=b_{2}/\hbar \sqrt{\frac{2 \gamma}{\rho \omega_{nl} M_{\mathrm{s}}}} \), where $b_2$ is the magnetoelastic coupling coefficient, $\hbar$ is the reduced Planck constant, and $M_s$ is the saturation magnetization \cite{guerreiro2015magonphonon}.

To identify the strong-coupling condition, we analyze the dispersion relation, as shown in Fig.~\ref{fig1}(b). For a permalloy  (\(\text{Ni}_{80}\text{Fe}_{20}\)) nanodisk ($d=20$ $\text{nm}$, $R=500$ $\text{nm}$), the magnon and phonon modes with \( l=1 \) intersect near 3.5 GHz, where the magnon and phonon may strongly couple. It is noted that higher-order OAM modes (\( l=2,3 \)) do not exhibit such crossing (see Fig. S1 \cite{sup}). This coupling facilitates efficient energy exchange, forming magnon polarons with hybrid magnon-phonon characteristics \cite{li2020hybrid}. The anticrossing gap \cite{hioki2022coherent}
\begin{equation}
\Delta f = \dfrac{b_2}{2\pi}\sqrt{\dfrac{2\gamma \omega_{l}}{C_{44} M_s}}
\label{eq:magnon--phonon_gap}
\end{equation}
quantifies the energy splitting due to the strong magnon-phonon coupling with \(C_{44}\) being the elastic stiffness. Using materials parameters (see below), one can estimate $\Delta f\approx 1.7$ GHz, defining the bandwidth of the strong coupling.

To describe the magnon-polaron dynamics, we apply the Bogoliubov transformation
\begin{equation}\label{Transformation}
  \left(
           \begin{array}{c}
             a_j \\
            b_j^{\dagger} \\
           \end{array}
         \right)=\left(
    \begin{array}{ccc}
      u_j & v_j \\
      v_j^* & u_j^*\\
    \end{array}
  \right)\left(
           \begin{array}{c}
             \alpha_j \\
             \beta_j^{\dagger} \\
           \end{array}
         \right),
\end{equation}
where  \(u_j\), \(v_j\) are coefficients for modes \(j = l, p,\) and \(q\) \cite{sup}. This diagonalizes the linear Hamiltonian ($H-H_{\mathrm{NL}}$). The nonlinear Hamiltonian then becomes
\begin{align}
H_{\text{NL}} &=g_{p}\left(u^{2} a_{g} \alpha_{l} \alpha_{p}^{\dagger}+v^{2} a_{g} \beta_{l}^{\dagger} \beta_{p}+ \text{H.c.}\right) \notag \\
&\quad+g_{q}\left(u^{2} a_{g}^{\dagger} \alpha_{l} \alpha_{q}^{\dagger}+v^{2} a_{g}^{\dagger} \beta_{l}^{\dagger} \beta_{q}+\text{H.c.}\right),\label{BV}
\end{align}
assuming \( u_l=u_p=u_q=u\), \( v_l=v_p=v_q=v\), valid when $\omega_g\ll\omega_l$ and $g_p\approx g_q\approx g$ \cite{sup}. Here, $u=\text{cosh}(\theta)$ and $v=\text{sinh}(\theta)$ with $\text{tanh}(2\theta)=g/\omega_{l}$. The nonlinear term \( H_{\text{NL}} \) now involves both the hybridized magnon-polaron modes (\( \alpha_j \), \( \beta_j \)) and the vortex core mode (\(a_g\)). It thus enables the nonlinearity transfer from magnons to phonons, as both the $\alpha$ and $\beta$ modes include the phonon component that is absent in Eq. \eqref{totalh}. This will facilitate new frequency components in the phonon spectrum, forming a PFC, as depicted in Fig.~\ref{fig1}(a).

Solving the Heisenberg equations, we derive the threshold  driving field \(h_c = {(\alpha^2 + \beta^2)\omega_{g}^2}/{g (u^4 + v^4)}\) \cite{sup}, above which the system enters the nonlinear regime, generating sum- and difference-frequency modes. Here, $\alpha$ and $\beta$ are damping parameters for the $\alpha$- and $\beta$-modes, respectively. The steady-state magnon-polaron populations are \cite{sup}
\begin{align}
|\langle \alpha_p \rangle| &= \frac{\sqrt{h_c (h - h_c)}}{\sqrt{2} \alpha (\omega_{g} + \omega_p)}, 
|\langle \alpha_q \rangle| = \frac{\sqrt{h_c (h - h_c)}}{\sqrt{2} \alpha (\omega_{q} - \omega_g)}, \label{Eq:p}\\
|\langle \beta_p \rangle| &= \frac{\sqrt{h_c (h - h_c)}}{\sqrt{2} \beta (\omega_{g} + \omega_p)}, 
|\langle \beta_q \rangle| = \frac{\sqrt{h_c (h - h_c)}}{\sqrt{2} \beta (\omega_{q} - \omega_g)}.\label{Eq:q}
\end{align}\label{polaron}

Below, we shall verify these analytical results by micromagnetic simulation using the MUMAX3 package \cite{vansteenkiste2014mumax3,vanderveken2021fdm}. In simulations, we adopt the saturation magnetization $M_s=800\,\text{kA}\text{/}\text m$, exchange stiffness $A=13\,\text{pJ}\text{/}\text m$, Gilbert damping constant $\alpha =0.008$, and elastic stiffness \(C_{44}=46\, \mathrm{GPa}\). These parameters align with recent experiments on permalloy nanodisks \cite{Vadym2024}, ensuring practical relevance. We set the magnetoelastic coupling constant $b_2 = 1 \times 10^7$ J/m$^3$ unless specified. The cell size is $5 \times 5 \times 5\,\mathrm{nm}^3$, with free boundary conditions for both the magnetic and elastic systems, initialized in a vortex state. 

\begin{figure}
    \centering
    \includegraphics[width=0.47\textwidth]{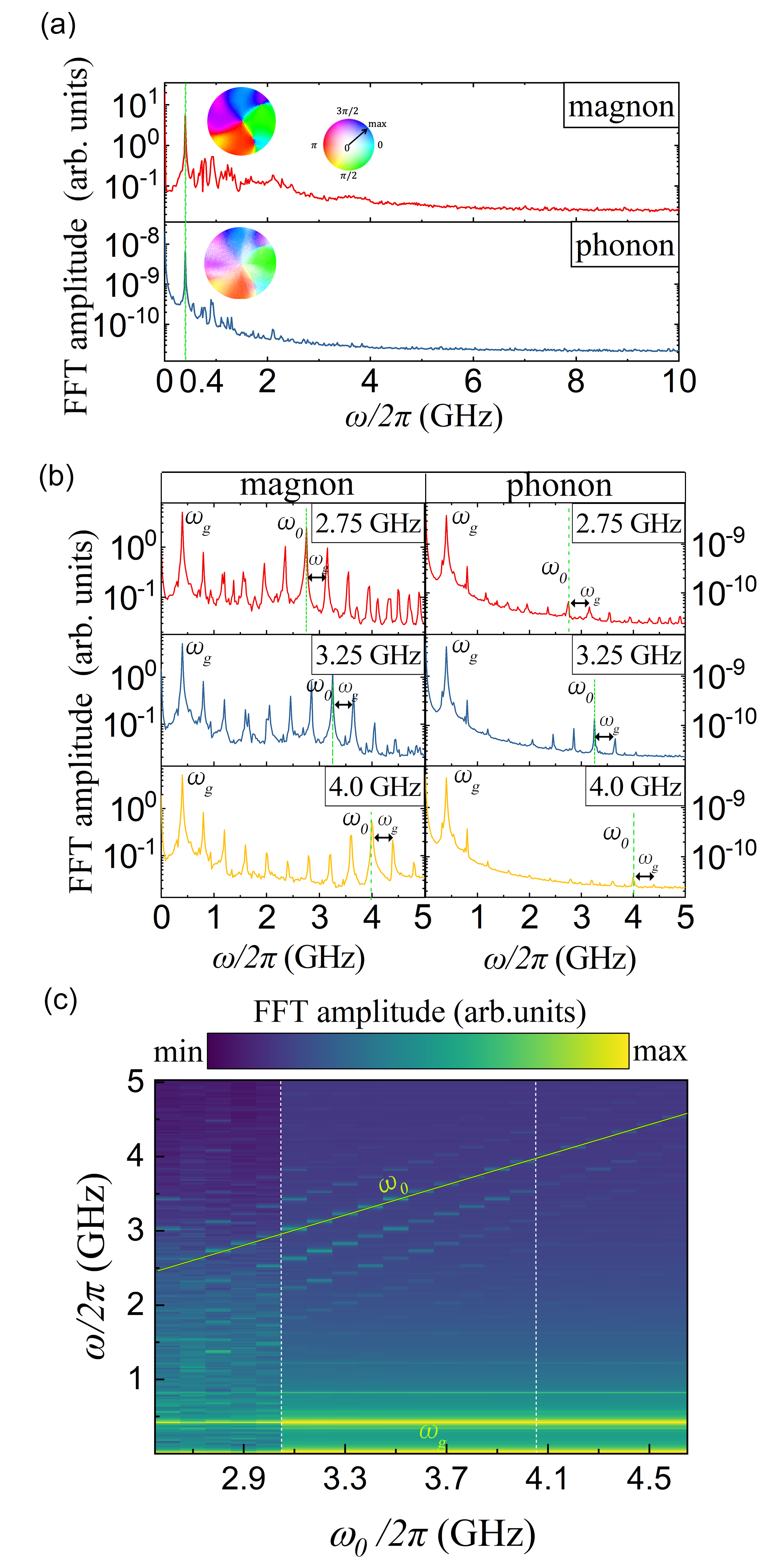}
    \caption{\label{fig2}
    (a) Magnon and phonon spectra of the vortex disk, showing large peaks at the vortex core gyrotropic frequency \(\omega_g\). Inset: spatial profiles of the gyrotropic vortex and lattice displacements at \(\omega_g\). 
    (b) Magnon and phonon spectra at driving frequencies \(\omega_0/2\pi=2.75, 3.25, 4.0 \) GHz. (c) Phonon response versus the driving frequency \(\omega_0\).}
\end{figure}

To characterize the coupled system's spectrum, we apply a sinc-function field \( \mathbf{h}(t) = h_0 \text{sinc}(\omega_c t)\hat{x}\) (\( \mu_0 h_0 = 1\,\text{mT} \), \(\omega_{c} / 2 \pi=10\,\text{GHz}\)) across the disk. Fast Fourier transform (FFT) of \(\delta m_{y}\) and \(\delta u_{y}\) [Fig.~\ref{fig2}(a)] reveals a main peak at 0.4 GHz, corresponding to the vortex core gyrotropic frequency. The 0.4 GHz mode appears in the lattice displacement spectrum as well, attributed to the vortex-like lattice deformation via the magnetoelastic coupling.

To study the excitation frequency effects, we apply an in-plane rotating field \( \mathbf{h}(t) = [h_0 \cos(\omega_0 t), h_0 \sin(\omega_0 t), 0] \) (\( \mu_0 h_0 = 1\,\text{mT} \)) to excite the magnon mode with \( l = +1 \) at \(\omega_0/2\pi=2.75\), \(3.25\) and \(4.0\, \mathrm{GHz}\). As shown in Fig.~\ref{fig2}(b), the magnon spectrum exhibits a frequency comb with the spacing of 0.4 GHz, matching the vortex core gyrotropic frequency. A prominent PFC emerges in the phonon spectrum only at 3.25 GHz, but not at 2.75 or 4.0 GHz outside the strong-coupling range (3.05--3.95 GHz) [region sandwiched by two dashed white lines in Fig.~\ref{fig2}(c)]. This confirms that a strong magnon-phonon coupling is essential for the nonlinearity transfer and PFC formation, consistent with our theoretical predictions.

\begin{figure} 
\centering
\includegraphics[width=0.48\textwidth]{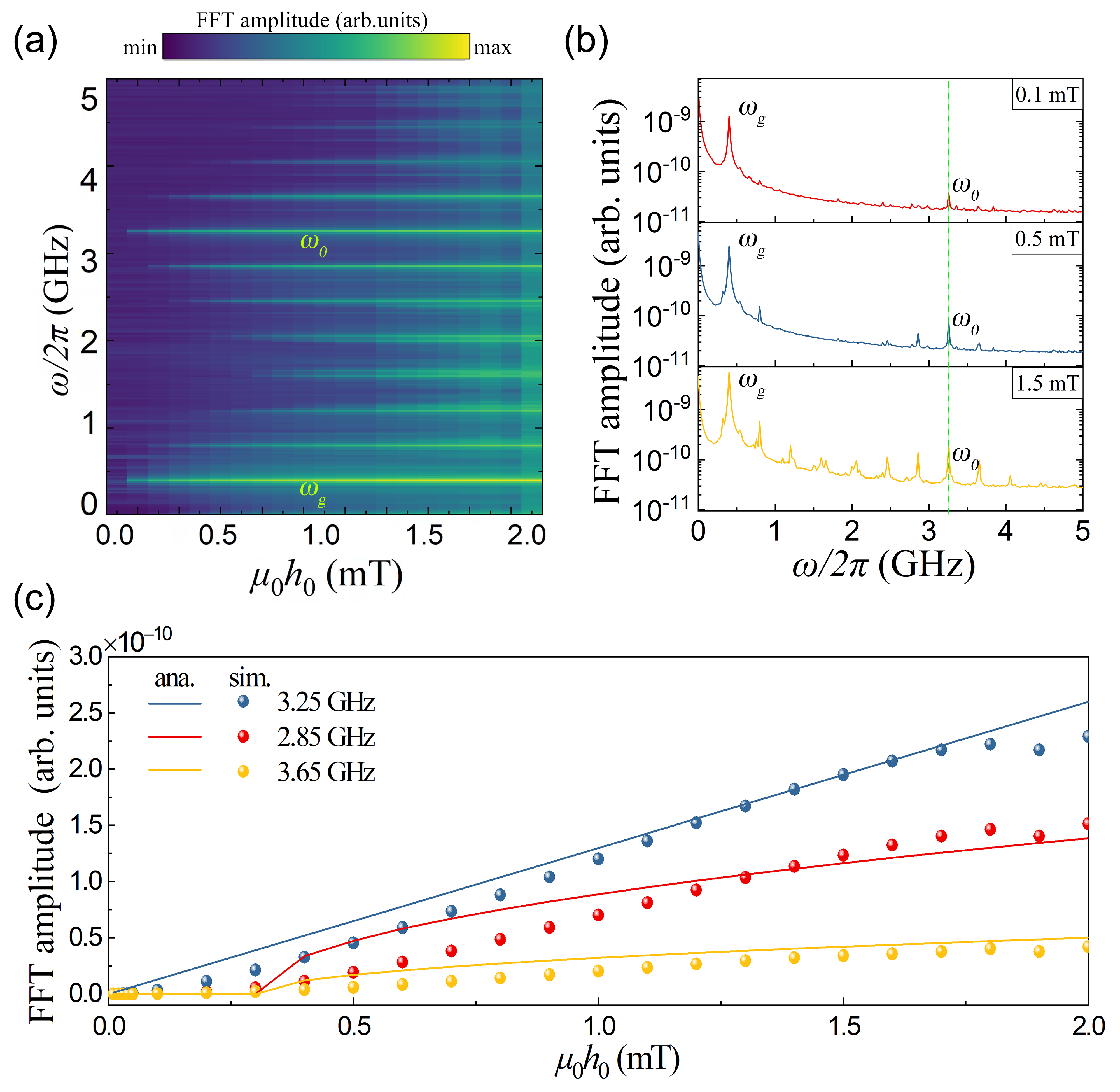} 
\caption{\label{fig3}
(a) Phonon response versus the driving field amplitude at \(\omega_0/2\pi=3.25\) GHz. (b) Phonon spectra at \( \mu_0 h_0 = 0.1\), \(0.5\), and \(  1.5\,\text{mT}\), respectively. (c) Amplitudes of peaks at \(\omega/2\pi=3.25\,\), \(2.85\) and \(3.65\, \mathrm{GHz}\) versus the field strength. Symbols: simulation results; curves: theoretical predictions.
} 
\end{figure}

To test the driving field's role, we apply an in-plane microwave field at \(\omega_0/2\pi=3.25\, \mathrm{GHz}\) with \(b_2 = 1.2 \times 10^7 \, \text{J/m}^3\). FFT of \(\delta u_y\) under varying field amplitudes [Fig.~\ref{fig3}(a)] reveals three regimes: \text{{(i)}} Below 0.3 mT, only the driving frequency (\( \omega_0 / 2\pi = 3. 25\, \text{GHz} \)) and the vortex core frequency (\( \omega_g / 2\pi = 0.4\, \text{GHz} \)) are excited [Fig.~\ref{fig3}(b)]. \text{{(ii)}} From 0.3 to 0.5 mT, side peaks at \(3.65=(3.25+0.4)\, \text{GHz}\) and \(2.85=(3.25-0.4)\, \text{GHz}\) emerge from the three-magnon confluence and splitting, respectively. \text{{(iii)}} Above 0.5 mT, a PFC forms with 0.4 GHz spacing. Figure~\ref{fig3}(c) shows the field-dependent amplitudes of the 3.25, 2.85, and 3.65 GHz peaks, with simulations (symbols) well matching theoretical predictions (curves), validating Eqs. \eqref{Eq:p} and \eqref{Eq:q}.

\begin{figure}
    \centering
    \includegraphics[width=0.5\textwidth]{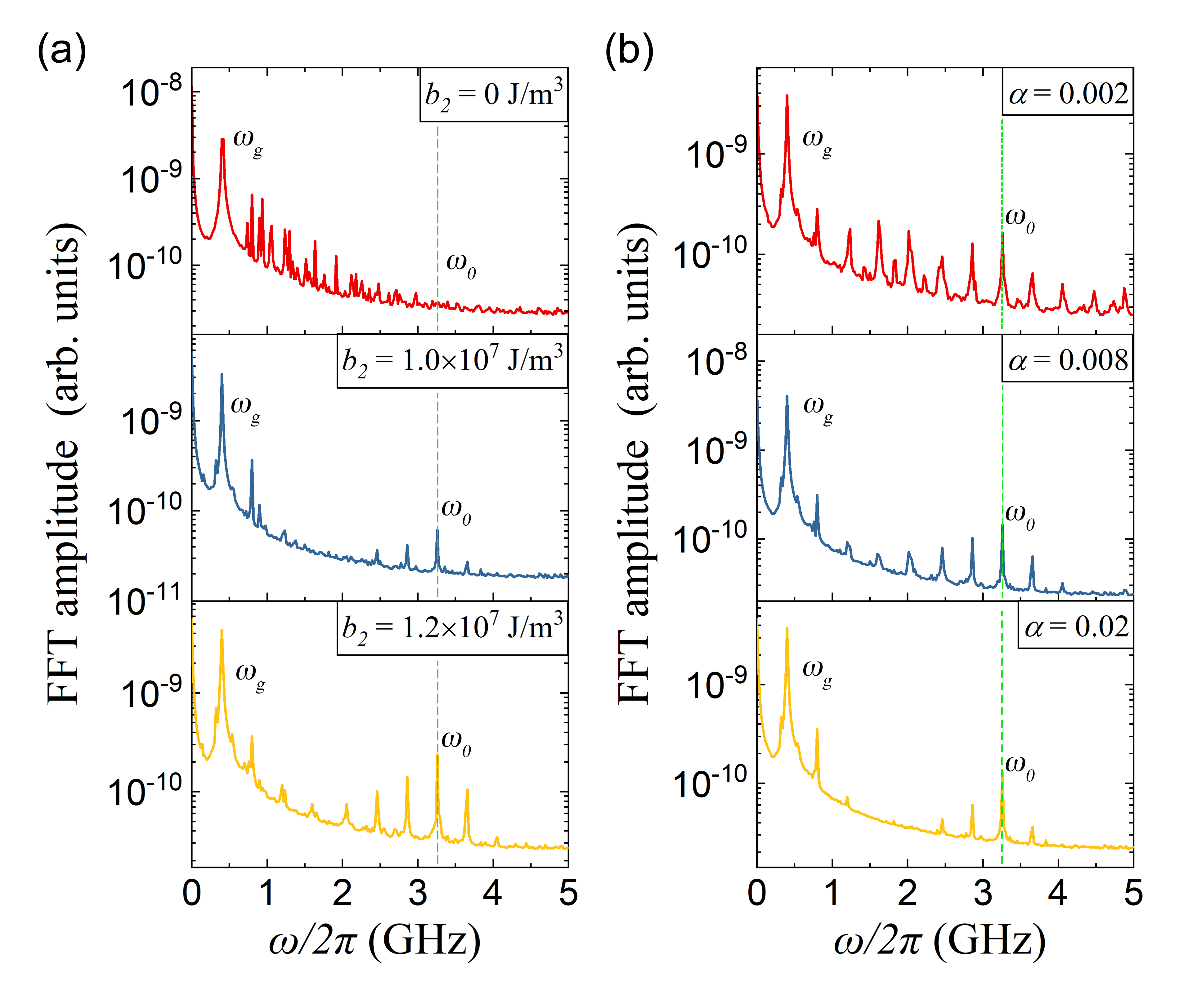}
    \caption{\label{fig4} 
  (a) Phonon spectra at magnetoelastic coupling strengths \(b_2 = 0\), \(1.0 \times 10^7\), and \(1.2 \times 10^7\) J/m\(^3\) with \(\alpha = 0.008\). (b) Phonon response at Gilbert damping coefficients \(\alpha = 0.002\), \(0.008\), and \(0.02\) with \(b_2 = 1.0 \times 10^7\) J/m\(^3\). In the simulations, we have fixed the driving frequency \(\omega_0/2\pi=3.25 \) GHz.
}
\end{figure}

The magnetoelastic coupling ($b_2$) and damping ($\alpha$) critically influence the PFC formation. Simulations by varying $b_2$ [Fig. \ref{fig4}(a)] show no PFC at $b_2 = 0$. As $b_2 $ increases, the comb's peak count and amplitude grow, with a broad, high-intensity comb at $b_2 = 1.2 \times 10^7$ J/m$^3$, due to the stronger magnon-phonon coupling ($g_{mp} \propto b_2$), enhancing the nonlinearity transfer \cite{sup}. It is also noted that the magnetoelastic coupling hardly changes the vortex gyration frequency. Conversely, increasing the damping strongly suppresses the comb [Fig. \ref{fig4}(b)]. At $\alpha = 0.002$, the comb is pronounced, but at $\alpha = 0.02$, high dissipation inhibits the nonlinear mode generation, thus limiting the PFC formation.

Thermal effects may play a significant role in the frequency comb generation. Elevated temperatures, often resulting from strong pump powers, can induce cyclic oscillations in magnon frequencies due to thermal shifts, leading to periodic instabilities in the comb structure. The increased thermal energy further disrupts the coherence between magnons and phonons, thereby affecting the stability and uniformity of the generated MFC \cite{xu2023magnonic}. To explore the thermal effect on the nonlinearity transfer, we conducted additional simulations at finite temperatures. Numerical results show that while the MFC is obviously suppressed by thermal fluctuations, the PFC can survive up to 300 \( \text{K} \) (see Fig. S2 \cite{sup}), demonstrating the robustness of the nonlinearity transfer mechanism.

To validate the predicted PFC structure experimentally, future studies could employ Brillouin light scattering or time-resolved magneto-optical Kerr effect measurements on permalloy nanodisks, directly probing the 0.4 GHz comb spacing. These methods, well-suited to resolving GHz-scale dynamics, would provide definitive evidence of the comb's spectral properties and their dependence on the vortex dynamics. The versatility of this mechanism may extend beyond the specific system studied here, as it can be readily adapted to other magnetic materials, such as YIG or cobalt-based nanostructures, or to alternative geometries like nanowires or thin films, or to other topological magnetization textures, e.g., skyrmions \cite{Yokouchi2020,Chen2023,Yang2024,Song2024} and bimerons \cite{Jani2021}. This adaptability establishes a robust platform for developing hybrid magnon-based systems that integrate mechanical and magnetic degrees of freedom. Exploring higher-order magnon or phonon modes could further enhance the stability and bandwidth of the combs, potentially yielding even finer frequency resolution or broader spectral coverage. Additionally, integrating these magnon-phonon systems with optical cavities or superconducting qubits could bridge the GHz-range PFCs with photonic or quantum platforms.

To conclude, we introduced a novel method to generate PFCs in purely linear elastic media. We showed that, by leveraging strong magnon-phonon coupling in a magnetic disk's vortex state, the nonlinearity effectively transfers from magnons to phonons, leading to the generation of PFC with comb spacing set by the vortex core's gyration frequency. Full micromagnetic simulations based on permalloy nanodisk parameters confirm robust GHz-range PFCs, overcoming traditional limits without nonlinear materials or intense drives. By redefining the boundaries of PFC generation, our result lays a foundation for next-generation devices that harness the interplay of spin and mechanical excitations across diverse fields in physics and engineering. Adaptable to various magnetic systems, it provides a universal strategy for frequency comb generation, advancing condensed matter physics and quantum information science.

\begin{acknowledgments}
The authors acknowledge financial support from the National Key R$\&$D Program under Contract No. 2022YFA1402802, the National Natural Science Foundation of China (Nos. 12104348, 12404085, 12434003, and 12374103), the Science and Technology Department of Hubei Province (No. 2024AFD012), and Sichuan Science and Technology Program (No. 2025NSFJQ0045).
\end{acknowledgments}

\end{document}